# NMR close to Mega-Bar Pressures

Thomas Meier[1*], Saiana Khandarkhaeva[1], Sylvain Petitgirard[1], Thomas Körber[2], Alexander Lauerer[3], Ernst Rössler[2], and Leonid Dubrovinsky[1]

1) Bayerisches Geoinstitut, Bayreuth University, Universitätsstraße 30, 95447 Bayreuth, Germany
2) Fakultät für Mathematik, Physik und Informatik, Experimentalphysik II, Bayreuth University, Universitätsstraße 30, 95447 Bayreuth, Germany
3) Institut für Materialwissenschaften, Hochschule Hof, Alfons-Goppel-Platz 1, 95028 Hof
*) Thomas.meier@uni-bayreuth.de

*Abstract*
The past 15 years have seen an astonishing increase in Nuclear Magnetic Resonance (NMR) sensitivity and accessible pressure range in high-pressure NMR experiments, owing to a series of new developments of NMR spectroscopy applied to the diamond anvil cell (DAC). Recently, with the application of electro-magnetic lenses, so-called Lenz lenses, in toroidal diamond indenter cells, pressures of up to 72 GPa with NMR spin sensitivities of about $10^{12}$ spin/Hz$^{1/2}$ has been achieved. Here, we describe the implementation of a refined NMR resonator structure using a pair of double stage Lenz lenses driven by a Helmholtz coil within a standard DAC, allowing to measure sample volumes as small as 100 pl prior to compression. With this set-up, pressures close to the mega-bar regime (1 Mbar = 100 GPa) could be realised repeatedly, with enhanced spin sensitivities of about $5\times10^{11}$ spin/Hz$^{1/2}$. The manufacturing and handling of these new NMR-DACs is relatively easy and straightforward, which will allow for further applications in physics, chemistry, or biochemistry.

*Introduction*
Varying thermodynamic conditions opens the possibility of accessing low-energy configurations, metastable or new states of matter, allowing the investigation of electronic or structural instabilities in solids. Thus, variation of pressure – that is directly reducing atomic or molecular distances – turned out to yield one of the most intriguing branches in condensed matter sciences [1], [2].
One of the most popular devices to generate high pressure is the diamond anvil cell (DAC) based on the Bridgman concept of a piston-cylinder press type. It was first introduced in the mid-fifties [3], and uses two diamond anvils to push together and compress a sample placed between their flattened faces. Since these first pioneering works many developments have been implemented such as the introduction of gaskets to confine the sample, pressure scale, pressure transmitting medium, laser heating. For the past two decades, it has now become a near routine to generate high pressure of 100 GPa in the laboratory, even reaching pressure found at the centre of the Earth [4], [5] and beyond [6], [7]. The success of the diamond anvil cells (DACs) resides in its astonishing variability in both design and field of application. To this end, *in-situ* spectroscopic methods in DACs are established, exploiting the diamonds' transparency to a broad range of wavelengths using lasers or X-Ray spectroscopy techniques covering most of the available pressure range. Other spectroscopic methods such as NMR or EPR, however, appear to be almost impossible to implement in DACs due to the following reasons: i) Sample cavities in DACs are typically tightly surrounded by both diamonds and a hard, metallic disc serving as a gasket. The gasket prevents the sample from leaving the region of the highest pressures, and provides so-called "massive support" to the stressed diamond anvils [8]. ii) Due to the necessarily small dimensions of the diamond anvils, available sample space is often much less than 5 nl before compression, which is further reduced when pressures exceeding several GPa are targeted. An application above 40 GPa, for example, requires an initial sample cavity of about 100 μm diameter and about 40 μm in height, amounting to about 350 pl. iii) Finally, the sample cavity is prone to plastic deformation under compression, leading to a volume reduction of the cavity of up to 50 % within a rather

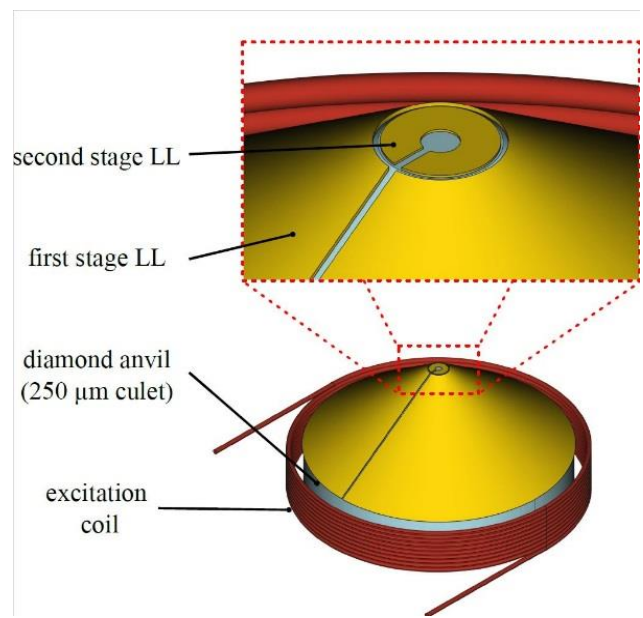

*Figure 1: Schematic design of the double stage Lenz lens (DSLL) resonator. Only one half of the complete assembly is shown. Both first and second stage LL are made from a 1 to 2 μm thick layer of copper deposited using PVD and cut into the depicted shape using a FIB. The driving coil, an 8-turn coil made from 100 μm copper wire is placed around the diamond anvil on the metallic anvil support (not shown). For further details, see text.*

small pressure range, depending on the choice of gasket material and pressure medium [9].

Thus, a successful implementation of NMR - or pulsed ESR for that matter – in diamond anvil cells, requires the implementation of resonators with suitable sizes and design. The first attempts have employed complex coil arrangements, either placed on the diamonds pavilion or over the whole diamond assembly, but did not allow measurements for pressures above 3 to 5 GPa [10]. A more promising solution has been the implementation of RF micro-coils directly into the high pressure sample chamber with measurements reaching pressures of up to 8 GPa [11], [12] and maximal pressures as high as 20 to 30 GPa [13], [14]. However, these minuscule micro-coils are extremely sensitive to the plastic deformation of the sample cavity, often exhibiting significant losses of $B_1$ field strength and subsequently NMR sensitivity by almost two orders of magnitude within a single pressure run [15], [16].

Recently, the application of electro-magnetic Lenz lenses in toroidal diamond indenter cells demonstrated that NMR at significantly higher pressures is not only feasible, but also comparatively easy to implement [17]. The basic principle of these magnetic flux tailoring devices is governed by Lenz's law of induction, hence the name Lenz lens (LL). Resonators using such LLs are typically driven by a bigger excitation coil directly connected to the NMR spectrometer. Following an RF pulse into the driving coil, the LL picks up the RF field via mutual inductance. The induced RF current is built up in the outer winding of the LL resonator and deposited in an inner region via a counter-winding, leading to a significant amplification of $B_1$ in a pre-defined volume. This basic idea of course makes LL resonators in DACs very attractive, as they can be used to focus the RF $B_1$ field where the high-pressure sample is located.

However, the latest design introduced by Meier et al. displayed some drawbacks. Its application in a DAC requires two diamond anvils with different culet diameter with, typically an 800 µm culet diamond on the cylinder side facing a 250 µm culet diamond on the piston side exerting the actual force. The main advantage of this technique is that the metallic rhenium gasket is buckled towards the much sharper piston anvil, leaving the space close to the 800 µm mostly untouched. This leaves enough room to place the RF excitation coil on the pavilion of the base anvil close to the culet, and thus to the 600 µm outer diameter LL used in these experiments. However, such anvil arrangements limit the accessible pressure range [18], with anvils damaged at a much smaller pressure range, often 60% below the standard capabilities of DAC experiments. Here, we introduce a new design and fabrication of RF resonators allowing for a further increase in maximal pressures and NMR sensitivity.

***Structure and preparation of the DSLL-resonator***

Figure 1 shows the principle design idea of the double-stage LL (DSLL) resonator.
The pressure cells equipped with these resonators were prepared as follows: After careful alignment of two 250 µm culet diamond anvils, a 250 µm thick rhenium disc was pre-indented to ~20 µm thickness.

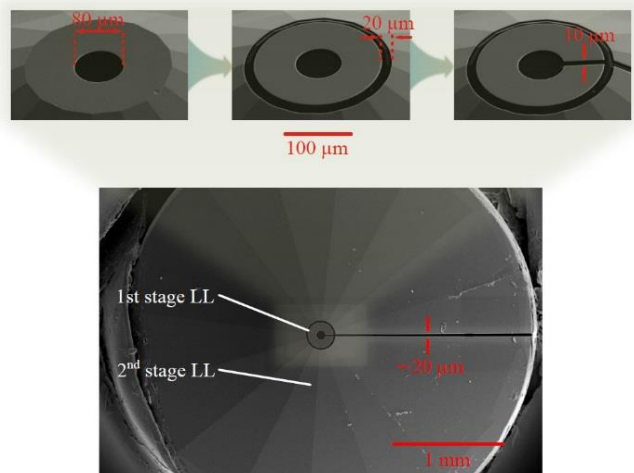

Figure 2: Representative SEM images of the DSLL resonator structure. The complete anvil can be seen on the left, incorporating both 1st and 2nd stage Lenz lenses. The slit in the 1st stage LL on the anvils pavilion is about 15 µm at its smallest point and increases a bit due to divergence of the gallium ion beam during cutting. The close up (right) shows the 2nd stage LL in detail. Bright spots on both photos are due to small dirt particles.

A ~80 µm sample hole was cut in the centre of the pre-indentation using an automated laser-drilling system at BGI.

PVD coating of the diamonds has been performed using a Dreva Arc 400 (manufacturer: VTD). An ultrapure copper target from Chempur (99.999 %) has been used. Argon (0.01 mbar) served as processing gas for plasma sputtering. The power of the Pinnacle magnetron power supply was set at 300 W. In order to achieve the required thickness of the copper layer (~ 2 µm) the duration of the coating process was set to 20 minutes. Using a focused ion beam (Scios Dual beam from FEI), the shape of the DSLL resonator was cut out from the almost homogeneous copper layer, using a 30 kV beam accelerator voltage and 65 nA gallium ion beam current.

Figure 2 shows SEM images during the DSLL resonator preparation on one of the diamond anvils. Additionally, the rhenium gaskets were covered with a 1 µm layer of $Al_2O_3$ on both sides providing electrical insulation between the lenses and the metallic gasket.

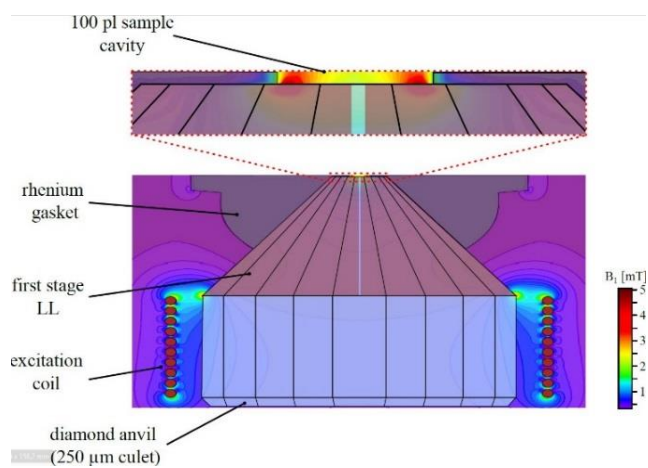

Figure 3: RF magnetic field simulation of the resonator set-up. As the assembly is symmetric, only the one half is shown.

The excitation coils were prepared from 100 µm PTFE insulated copper wire and consisted of 8 turns with a diameter of 4 mm. Both coils were placed on the backing plates of the diamonds, fully enclosing the anvils. Subsequently, the prepared anvils were aligned again in the DACs, and the cells were loaded with distilled water and slightly closed to prevent water leakage.

After the cells were closed, both coils were connected in order to form a Helmholtz arrangement. The loaded and pressurised cells were then mounted on a home built NMR probe for standard wide-bore magnets. Analysing the return-loss spectrum of the resonator at 400 MHz, a quality factor of the resonance circuit of about 40 was found.

All measurements were conducted at a magnetic field of 9 T. The actual pressure in the sample cavity was monitored using the first derivative of the pressure dependent first order Raman vibron mode taken at the centre of the diamonds' culets [19],[20].

### Analysis of NMR performance and stability at 90 GPa

As a first step, numerical simulations of the RF $B_1$ field of the DSLL resonator have been conducted using the FEMM software package[21]. Figure 3 shows the simulation results for one half of the full assembly, taking into account all parts of the high pressure resonator, including both LLs placed on one anvil, one half of the Helmholtz coil, and the electrically insulated rhenium gasket. As can be seen, application of the DSLL resonator leads to a focusing effect of the $B_1$ field strength due to significant reduction in the final diameter of the resonator. In the 100 pl sample cavity, the $B_1$ field varies between 2.5 mT at the centre up to 5 mT at the inner circumference of the 2$^{nd}$ stage LLs at 40 µm from the centre. A solitary application of the Helmholtz coil would only amount to magnetic fields of about 0.8 to 1.1 mT at the high pressure centre (simulations are not shown).

Using RF nutation experiments at 90 GPa, an optimal 90° pulse length of $t_{\pi/2}=2.5$ µs at 10 W average pulse power was found, leading to an actual $B_1$ field strength of $B_1 = \pi/(\gamma_n t_{\pi/2}) = 2.3$ mT, which is in excellent agreement with the found values from the numerical simulation.

Estimating the sensitivity of this new resonator, the common definition of the time-domain limit of detection ($LOD_t$) as the minimal necessary number of spins, resonating in a 1 Hz bandwidth and providing a single-shot SNR of unity in the time domain, is used. Figure 4 shows data from a solid echo train recorded at 90 GPa of compressed $H_2O$, well within the stability field of ice X [22]. At pulse separations of 20 µs, a single shot SNR of 39 recorded with a bandwidth of 2 MHz could be achieved. Considering a number of approximately $3\cdot10^{16}$ $^1H$ nuclei present within the 100 pl cavity, a $LOD_t$ of about $5\times10^{11}$ spin/Hz$^{1/2}$ was found. Additional investigations of LODt over the entire pressure range from 8 GPa to 90 GPa did not display any deviations in the spin sensitivities exceeding 2% of this value. A thorough analysis the of behaviour of high pressure ices VII and X will be presented elsewhere [23].

Another important point for the characterisation of a high pressure resonator is its ability to withstand high mechanical stresses and deformation under load. Figure 5 depicts both recorded Raman spectra at the diamond edge, as well as X-ray transmission measurements performed by scanning the X-ray intensity along a fixed axis across the high pressure region of the DAC. As can be seen, the diamond anvils become highly stressed at increasing pressures, leading to a significant line broadening of the Raman vibron mode. The transmission X-ray profile shows both diamond anvils clearly cupped at 90 GPa – as can be seen by the increased X-ray absorption in the region between the sample cavity and the culet, which is at about 25-125 µm. In fact, the equality of X-ray intensity at the culet edge and within the sample cavity signifies the diamonds almost touching: further increase of pressure would not be possible without diamond breakage. Interestingly, both the cavity diameter as well as the inner diameter of the 2$^{nd}$ stage LLs at the diamond culet did not decrease significantly from their initial values.

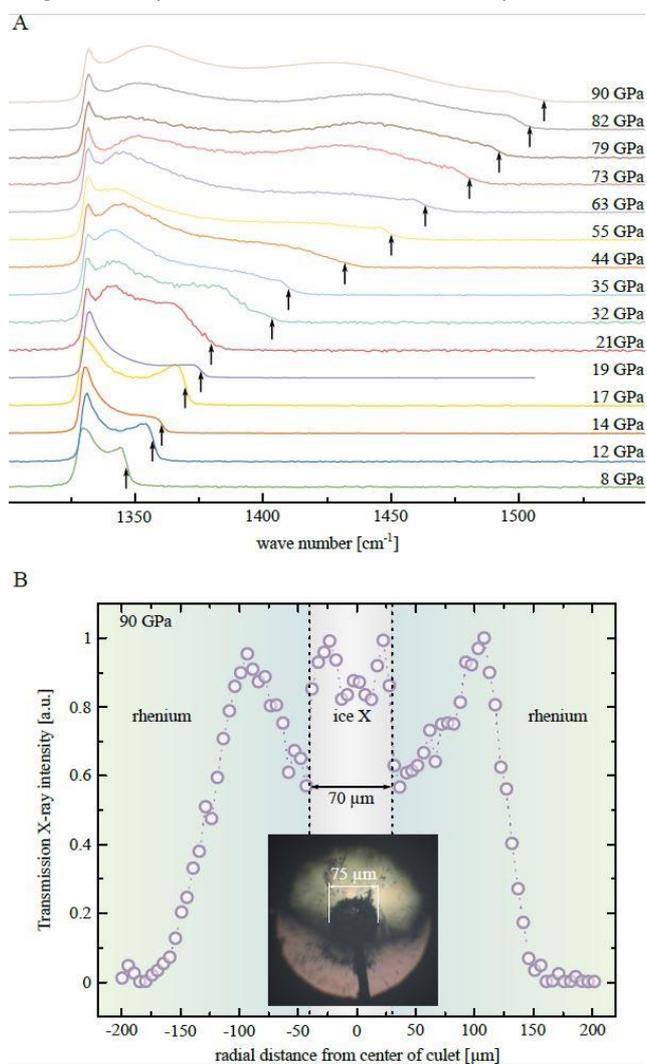

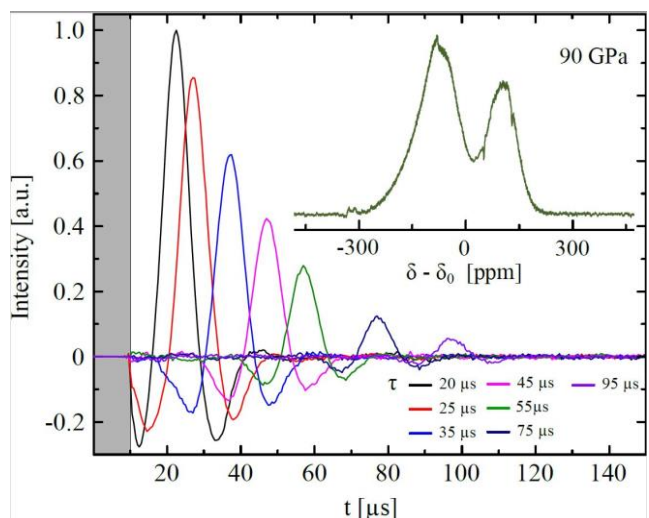

Figure 4: Recorded π/2-τ- π/2 solid echo train of high pressure ice X at 90 GPa. The spectrometer was blanked off for 4.6 µs after the second 90° pulse (grey area). The inset shows a respective Fourier transform NMR spectrum

Figure 5: A) Raman spectra acquired at the diamond edge at the center of the high pressure region in the DAC. Black arrows show the position of the minima of the spectras' first derivative which was used for pressure determination. B) X-ray absorption profile along a fixed axis across the inner part of the DAC. Inset: photograph of a 2$^{nd}$ stage LL from within a DAC pressurized to 90 GPa.

Both 2$^{nd}$ stage LLs where found to be intact even at 90 GPa pressure at the centre of the DAC, as can be seen in the photograph in figure 4B. This is in stark contrast to the previously used LLs made from 5 µm thick Au-foil, all of which displayed pressure dependent deformation at pressures exceeding about 19 GPa.

### *Discussion and Conclusions*

The presented DSLL-resonator for high-pressure NMR applications exhibits several significant advances compared to previous set-ups. First, these novel resonators allow for a stable and safe use in a standard DAC equipped with two diamonds of identical culet diameters. As electro-magnetic coupling between excitation coils, 1$^{st}$ stage, and 2$^{nd}$ stage LLs works sufficiently well, it may be applicable for even smaller culet faces, opening the possibility for NMR at even high pressures. This conjecture might be confirmed, following some geometric reasoning:

Preparation of the gasket typically follows some empirical rules found to maximise the DAC's stability under load, namely that the diameter of the sample chamber should be around 1/3 x d, whereas the pre-indentation is usually as flat as 1/6 x d, where d is the culet diameter of the diamond anvils. This sets some upper limits for the sample volume $V_0$ prior to compression, i.e. $V_0 \approx 1.5 \cdot 10^{-2} \cdot d^3$.

Empirical maximal pressures in a standard DAC for non-NMR applications was found to be proportional to $\sim d^{-2}$ up to about 1.5 Mbar, with a flattening-out of this behaviour to roughly a $\sim d^{-1/2}$ dependence above this pressure. The inset in figure 6 illustrates these dependencies.

Obviously, the use of smaller diamond anvil culets will inevitably reduce both diameter and height of the sample cavity. In contrast, the diameter of the counter winding as well as the separation between both 2$^{nd}$ stage LLs will also be reduced, leading to a further increase in $B_1$. Therefore, the inevitable loss in SNR due to reduced amount of sample for smaller $V_0$ and maximal pressures, will be compensated by an increase in LOD$_t$ due to this proximity boost.

The deduced spin sensitivities of about $10^{11}$ spin/Hz$^{1/2}$ mark a major advancement in this application field, which is illustrated in the main frame of figure 6. Here, we summarized extracted data on LOD$_t$ from the majority of all known high-pressure NMR set-ups, and compared it to their pressure stability.

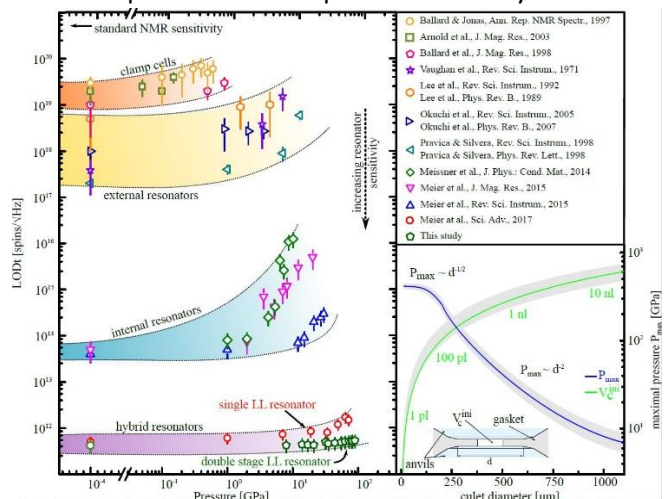

*Figure 6: Main frame: NMR limits of detection at increasing pressures for several different high pressure NMR set-ups. For more information, see text. Right inset: dependence of sample volume V as well as the maximal achievable pressure for different diamond culet diameters d.*

One of the most widely used applications for high pressure NMR is bio-chemistry, in particular protein-folding dynamics [24]–[30], using clamp cells which allow for a significantly lower pressure compared to standard DACs. LOD$_t$ values in these cases often range between 1 - 3·10$^{19}$ spin/Hz$^{1/2}$ with maximal pressures of about 1 GPa. Within DACs, however, three distinct groups can be identified from figure 6: external, integral, and hybrid resonators.

External resonators encompass all set-ups introduced from the end of the 1980s to 1998, that is resonators which are either placed solely on the pavilion of the anvils [31]–[35] or single turn cover inductors on top of the rhenium gaskets[36], exhibiting LOD$_t$ of about 8x10$^{18}$ to 2x10$^{17}$ spin/Hz$^{1/2}$ at pressures as high as 13 GPa[37]. Internal resonators, on the other hand, comprise micro-coils placed directly into the sample chamber with sensitivities from about 1x10$^{16}$ spin/Hz$^{1/2}$ to 6·10$^{13}$ spin/Hz$^{1/2}$ depending on the degree of coil deformation within each pressure run.

The group of hybrid resonators ("hybrid" because both external and internal resonators are used together), comprise the recently introduced single LL in a toroidal diamond indenter cell as well as the DSLL-resonator in a DAC, introduced here. From figure 6 it is obvious that LOD$_t$ is not only increased by almost eight orders of magnitude compared to clamp cell NMR sensitivities but they were also found to be exceedingly stable over the entire pressure run, originating in the flat, almost two-dimensional design of the LLs used.

Finally, the preparation process of these novel resonators could be simplified and developed further up to a point enabling a larger number of NMR laboratories to use high pressure NMR as an convenient everyday investigative tool.

### Acknowledgements

We thank Nobuyoshi Miyajima and Katharina Marquardt for provision of the FIB, and help with the ion milling (grant number: INST 90/315-1 FUGG). We are also very thankful for the help of Sven Linhardt and Stefan Übelhack for manufacturing the NMR probe and pressure cell components. The authors, T.M., S.P., and L.D., were funded by the Bavarian Geoinstitute through the Free State of Bavaria. S.K. was funded through the German Research Society (DU-393/13-1). PVD has been performed at Hof University of Applied Science: Prof. Dr. Jörg Krumeich, Dipl.-Ing. Katrin Huget and M. Eng. Stephan Paulack are gratefully acknowledged.